\documentclass[a4,12pt]{article}

\usepackage{a4wide}
\usepackage{amssymb}
\usepackage{amsmath}
\usepackage{graphicx}
\usepackage{xspace}
\usepackage{epsfig}
\usepackage{url}
\usepackage{cite}

\newcommand{\order}[1]{\mathcal{O}\left(#1\right)}
\newcommand{\ie}{i.e.\ }
\newcommand{\eg}{e.g.\ }
\newcommand{\cG}{{\cal G}}
\newcommand{\acknowledgements}{\section*{Acknowledgements}}
\renewcommand{\maketitle}{\vspace{2em}}
\newcommand{\VoronoiWidth}{0.4\textwidth}
\newcommand{\TimingsWidth}{0.7\textwidth}

\begin{document}

%%%%%%%%%%%%%%%%%%%%%%%%%%%%%%%%%%%%%%%%%%%%%%%%%%%%%%%%%%%%%%%%%%%%%%%
\titlepage
\begin{flushright}
hep-ph/0512210 \\
LPTHE--05--32\\
December 2005\\
revised August 2006
\end{flushright}

\vspace*{0.3in}
\begin{center}
{\Large \textbf{\textsf{
Dispelling the $\boldsymbol{N^3}$ myth for the $\boldsymbol{k_t}$ jet-finder
}}}\\
\vspace*{0.4in}
Matteo~Cacciari and
Gavin~P.~Salam \\
{\small
\vspace*{0.5cm}
LPTHE, Universities of Paris VI and VII and CNRS,
}\\
\vskip 2mm
\end{center}
%\vspace*{1cm}
%\centerline{(\today)}

\begin{abstract}
  At high-energy colliders, jets of hadrons are the observable
  counterparts of the perturbative concepts of quarks and gluons.
  Good procedures for identifying jets are central to experimental
  analyses and comparisons with theory. The $k_t$ family of successive
  recombination jet finders has been widely advocated because of its
  conceptual simplicity and flexibility and its unique ability to
  approximately reconstruct the partonic branching sequence in an
  event.
  Until now however, it had been believed that for an ensemble of $N$
  particles the algorithmic complexity of the $k_t$ jet finder scaled
  as $N^3$, a severe issue in the high multiplicity environments of
  LHC and heavy-ion colliders.
  We here show that the computationally complex part of $k_t$
  jet-clustering can be reduced to two-dimensional nearest neighbour
  location for a dynamic set of points. Borrowing techniques developed
  for this extensively studied problem in computational geometry,
  $k_t$ jet-finding can then be performed in $N \ln N$ time.
  Code based on these ideas is found to run faster than all other jet
  finders in current use. 
\end{abstract}

\maketitle

%\vspace{0.5cm}
%\vspace{0.5cm}
%%%%%%%%%%%%%%%%%%%%%%%%%%%%%%%%%%%%%%%%%%%%%%%%%%%%%%%%%%%%%%%%%%%%%%%

%\begin{multicols}{2}

%======================================================================
\section{Introduction}
\label{sec:introduction}

Partons (quarks and gluons), are the concepts that are central to
discussions of the QCD aspects of high-energy collisions such as those
at the Fermilab Tevatron and the future Large Hadron Collider (LHC) at
CERN. Quarks and gluons, however, are not observable, and in their
place one sees \emph{jets,} collimated bunches of high-energy hadrons
which are the result of the fragmentation and hadronisation of the
original hard (high-energy) partons.  Today's limited understanding of
non-perturbative QCD is such that it is not currently possible to predict
the exact patterns of hadrons produced. Instead one makes predictions
in terms of quarks and gluons and relates these to observations in
terms of hadron jets.

Naively, jets are easily identified --- one simply searches for
bunches of collimated hadrons. However, to carry out accurate
comparisons between parton-level predictions and hadron-level
observations one needs a well-defined `jet-finding'
procedure.  The jet-finder is applied both to perturbatively predicted
partonic configurations and to observed hadronic configurations and
one then directly compares distributions for the predicted partonic
jets and the observed hadronic jets. 
Though partonic and hadronic jets are not %strictly speaking
equivalent, there is strong evidence
(theoretical~\cite{Sterman:1977wj} and
experimental~\cite{ExpJetEvidence}) that the comparison can be
performed with controlled accuracy.

Insofar as jet-finding is an approximate attempt to invert the quantum
mechanical processes of QCD branching and hadronisation, it is not a
unique procedure.  Various kinds of jet-finders have been proposed,
among them cone-type \cite{Sterman:1977wj,Cone} and
sequential-clustering \cite{Bartel:1986ua,Kt,ARCLUS,CamAachen}
jet-finders (for alternatives, see
\cite{JetEnergyFlow,Tkachov,DeterministicAnnealing,Chekanov:2005cq}).

Cone jet-finders are the most frequently used at the Tevatron. They are
based on identifying energy-flow into cones in (pseudo)rapidity
$\eta=-\ln \tan \theta/2$ and azimuth $\phi$, together with various steps of
iteration, merging and splitting of the cones to obtain the final
jets. Cone jet-finders tend to be rather complex, different
experiments have used different variants (some of them infrared unsafe),
and it is often difficult to know exactly which jet-finder to use in
theoretical comparisons.

In contrast, the cluster-type jet-finders, generally based on successive
pair-wise recombination of particles, have simple definitions and are
all infrared safe (for reviews see
\cite{Moretti:1998qx,ChekanovReview}). We shall
focus here on the most widely used of them, the
$k_t$ jet-finder~\cite{Kt}, defined below. Among its physics
advantages are  (a) that it
purposely mimics a walk backwards through the QCD branching sequence,
which means that reconstructed jets naturally collect most of the
particles radiated from an original hard parton, giving better
particle mass measurements \cite{KtVersusCone,KtVersusConeBis},
general kinematic reconstruction~\cite{LesHouches} and
gaps-between-jets identification \cite{KtGaps} (of relevance
to Higgs searches); and (b) it allows one
to decompose a jet into constituent subjets, which is useful for
identifying decay products of fast-moving heavy particles (see \eg
\cite{KtSubJetAnalysis}) and various QCD studies. This has led to the
widespread adoption of the $k_t$ jet-finder in the LEP ($e^+e^-$
collisions) and HERA ($ep$) communities.

Despite its advantages, $k_t$ clustering has so far seen only limited
study \cite{D0kt,D0KtSubJets,CDFkt} at the Tevatron. The reasons for this are not
entirely clear. One known drawback of the $k_t$ jet finder for
high-multiplicity hadron-collider environments is its apparent
algorithmic slowness: to cluster $N$ particles into jets requires
$\order{N^3}$ operations in current implementations
\cite{KtImplementation}.  For a typical 
event at the upcoming LHC, with an expected
multiplicity of $N = \order{2000}$, this translates into a clustering
time of $\order{10~s}$ of CPU time on a modern
$\order{3~{\mathrm{GHz}}}$ processor; this is considerable given that
the clustering has to be repeated for millions of events.  For a
typical heavy-ion event at LHC, where $N = \order{50000}$, the
clustering time would grow to an unsustainable $\order{10^5~s}$, i.e.
more than one day! 
Even at the Tevatron, where the multiplicity is quite modest, the fact
that noise may cause the number of active calorimeter cells to be far
larger than the number of particles has led to the use of a complex
(and physically questionable) preclustering procedure prior to running
the $k_t$ jet finder, so as to reduce the effective value of $N$ to
something that is manageable~\cite{D0KtSubJets}.

The slowness of the $k_t$ jet-finder has been one of the motivating
factors behind proposals for alternative jet-finders
\cite{Tkachov,DeterministicAnnealing}. Here we will show that the
$k_t$ jet-finder can in fact be formulated in an algorithmically fast
($N \ln N$) manner. A {\tt C++} implementation of this (and a related $N^2$)
algorithm\footnote 
{`Jet-algorithm' is often used in the literature to refer to the
  choice of the rules for finding a jet; here instead `algorithm'
  refers to the translation of a given set of jet-finding rules into
  explicit steps on a computer.}
will be shown to run orders of magnitude faster than currently
available implementations, making it feasible (and easy) to use the $k_t$ jet
finder for efficiently studying high-multiplicity events.

\section{The $k_t$ jet-finder}
\label{sec:kt}

The $k_t$ jet finder, in the longitudinally invariant formulation
suitable for hadron colliders, is defined as follows. 
\begin{itemize}
\item[1.] For each pair of particles $i$, $j$ work out the $k_t$
  distance $d_{ij} = \min(k_{ti}^2,{k_{tj}^2}) R_{ij}^2$ with
  $R_{ij}^2 = (\eta_i-\eta_j)^2 + (\phi_i-\phi_j)^2$, where $k_{ti}$,
  $\eta_i$ and $\phi_i$ are the transverse momentum, rapidity and
  azimuth of particle $i$; for each parton $i$ also work out the beam
  distance $d_{iB} = k_{ti}^2$.
\item[2.] Find the minimum $d_{\min}$ of all the $d_{ij},d_{iB}$. If
  $d_{\min}$ is a $d_{ij}$ merge particles $i$ and $j$ into a single
  particle, summing their four-momenta (alternative recombination
  schemes are possible); if it is a $d_{iB}$ then declare particle $i$
  to be a final jet and remove it from the list.
\item[3.] Repeat from step 1 until no particles are left.
\end{itemize}
There exist extensions of this basic procedure, (a) where $d_{ij}$ is
rescaled relative to $d_{iB}$ by a user-chosen factor $1/R^2 \sim 1$
or (b) where clustering is stopped when all $d_{ij},d_{iB}$ are above
a jet resolution threshold $d_{\mathrm cut}$. We here consider only the
simplest version, as given above, but the arguments below are identical for
the general case.

Now we reconsider the above procedure, making explicit the
computational overheads of the various steps as implemented in
standard jet finding codes \cite{KtImplementation}.
\begin{enumerate}
\item Given the initial set of particles, construct a table of all the
  $d_{ij}$, $d_{iB}$.\\
  \mbox{ }\hfill
  [\emph{$\order{N^2}$ operations, done once%
  %   ,\\ \mbox{ }\hfill
  %  using $\order{N^2}$ storage%
  }]
\item Scan the table to find the minimal value $d_{\min}$ of the
  $d_{ij}$, $d_{iB}$.  \\\mbox{ }\hfill [\emph{$\order{N^2}$
    operations, done $N$ times}]
\item Merge or remove the particles corresponding to $d_{\min}$ as
  appropriate.\\\mbox{ }\hfill
  [\emph{$\order{1}$ operations, done $N$ times}]
\item Update the table of $d_{ij}$, $d_{iB}$ to take into account the
  merging or removal, and if any particles are left go to step 2.\\
  \mbox{ }\hfill [\emph{$\order{N}$ operations, done $N$ times}]
\end{enumerate}
Step 2 dominates, requiring $\order{N^2\times N = N^3}$ operations.\footnote{
 One notes also
the storage requirement in step 1 of $4N^2 + \order{N}$ bytes (double
precision), which is manageable for $N=1000$ but becomes an issue
in heavy-ion environments with up to $50000$
particles. At the (substantial) expense of recalculating the
  $\order{N^2}$ $d_{ij}$ at each iteration, the storage issue can
  be eliminated.}

\section{The {\tt FastJet} Algorithm}
\label{sec:fastjet}

To obtain a better algorithm we isolate the geometrical aspects of the
problem, with the help of the following observation.

\textbf{Lemma:} If $i$, $j$ form the smallest $d_{ij}$, and $k_{ti} <
k_{tj}$, then $R_{ij} < R_{i\ell}$ for all $\ell \neq j$, \ie $j$ is
the geometrical nearest neighbour of particle $i$.

\textbf{Proof:} Suppose the Lemma is wrong and that there exists a particle
$\ell$ such that $R_{i\ell} \le R_{ij}$: then $d_{i\ell} =
\min(k_{ti}^2,k_{t\ell}^2) R_{i\ell}^2$ and since
$\min(k_{ti}^2,k_{t\ell}^2) \le k_{ti}^2$, we have that 
%%%%%%%%% $d_{i\ell} \le d_{ik}$,       d_{ij}, right?!?!?!?!?!? 
$d_{i\ell} \le d_{ij}$, 
in contradiction with the statement that $i$ and $j$ have the
smallest $d_{ij}$.

This means that if we can identify each particle's
geometrical nearest
neighbour (in terms of the geometrical $R_{ij}$ distance), then we
need not construct a size-$N^2$ table of 
$d_{ij} = \min(k_{ti}^2,{k_{tj}^2}) R_{ij}^2$, but only the
size-$N$ array, $d_{i{\cG\!}_i}$, where $\cG_i$ is $i$'s $\cG$eometrical 
nearest neighbour\footnote{We shall drop `geometrical' in the
following, speaking simply of a `nearest neighbour'.}.
We can therefore write the following algorithm:
\begin{enumerate}
\item For each particle $i$ establish its nearest neighbour $\cG_i$ and
  construct the arrays of the $d_{i{\cG\!}_i}$ and $d_{iB}$.
\item Find the minimal value $d_{\min}$ of the $d_{i{\cG\!}_i}$, $d_{iB}$.
\item Merge or remove the particles corresponding to $d_{\min}$ as
  appropriate.
\item Identify which particles' nearest neighbours have changed and
  update the arrays of $d_{i{\cG\!}_i}$ and $d_{iB}$. If any particles are
  left go to step 2.
\end{enumerate}
This already reduces the problem to one of
complexity $N^2$: for each particle we can find its  
nearest neighbour
by scanning through all $\order{N}$ other particles [$\order{N^2}$
operations]; calculating the $d_{i{\cG\!}_i}$, $d_{iB}$ requires $\order{N}$
operations; scanning through the $d_{i{\cG\!}_i}$, $d_{iB}$ to find the
minimal value $d_{\min}$ takes $\order{N}$ operations [to be repeated
$N$ times]; and after a merging or removal, updating the nearest
neighbour information will require $\order{N}$ operations [to be
repeated $N$ times].\footnote{This last point is not strictly speaking
  trivial: when particle $i$ is removed or merged we have to update
  the nearest neighbour information for all particles that previously
  had $i$ as their nearest neighbour --- fortunately one can show
  that on average, the number of particles that had $i$
  as a nearest neighbour is of $\order{1}$.
  One also needs to establish if any particles acquire the newly
  created particle $\ell$ as their nearest neighbour --- this can be
  done in $\order{N}$ time by comparing each particle's current
  nearest neighbour distance with its distance to $\ell$.
} %

\begin{figure}[t!]
\begin{center}
%\vspace{-1.5cm}
%\includegraphics[width=\columnwidth]{voronoi.ps}
\includegraphics[width=\VoronoiWidth]{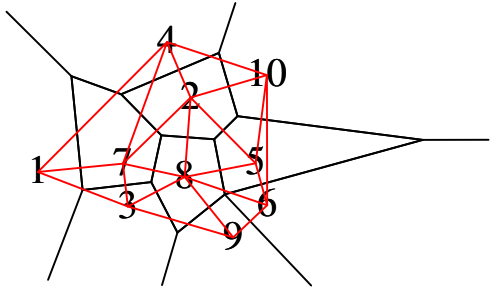}
%\vspace{-1.5cm}
\caption{\label{fig:voronoi} \small The Voronoi diagram for
ten random points. The Delaunay triangulation (red) connecting the ten
points is also shown. In this example the points 1, 4, 2, 8 and 3 are the
`Voronoi' neighbours of 7, and 3 is its nearest neighbour.}
\end{center}
\end{figure}

We note, though, that three steps of this algorithm ---
initial nearest neighbour identification, finding $d_{\min}$ at each
iteration, and updating the nearest neighbour information at each
iteration --- bear close resemblance to problems
studied in the computer science literature and for which efficient
solutions are known:
\begin{itemize}
\item Given an ensemble of vertices in a plane (specified by the $\eta_i$ and
  $\phi_i$ of the particles), to find the nearest
  neighbour of each vertex one can use a structure known as a Voronoi
  diagram \cite{WikipediaRefs} or its dual, a Delaunay triangulation.
  The Voronoi diagram divides the plane into cells (one per vertex),
  such that every point in the cell surrounding a vertex $i$ has $i$
  as its nearest vertex.
  The structure is useful for nearest-neighbour location because the
  vertex $\cG_i$ nearest to vertex to $i$ is always in one of the
  (few, \ie $\order{1}$) cells that share an edge with the cell of
  vertex $i$.  An example is shown in figure~\ref{fig:voronoi}.
  Voronoi diagrams for $N$ points can be constructed with $\order{N
    \ln N}$ operations (see \eg \cite{Fortune}), and the nearest neighbour
  identification for all $N$ points can then be performed with a
  further $\order{N}$ operations.
\item Dynamic insertion and removal of a point in the Voronoi diagram,
  and corresponding updating of all nearest neighbour information, can
  be performed with $\order{\ln N}$ operations \cite{DelaunayDeletion}
  (to be repeated $N$ times).
\item The array of $d_{i{\cG\!}_i}$ changes only by $\order{1}$
  entries per iteration. Therefore one can represent it with a binary
  tree structure, whose construction requires $\order{N \ln N}$ operations and
  in which finding the minimal value, insertion and removal are all
  guaranteed to require at most $\order{\ln N}$. The binary tree is
  constructed once at start-up, and there are then $\order{N}$ updates
  and searches for the minimum, leading to a total of $\order{N \ln
    N}$ operations.
\end{itemize}
Therefore both the geometrical and minimum-finding aspects of the
$k_t$ jet-finder can be related to known problems whose solutions
require $\order{N \ln N}$ operations.

\section{Timings}
\label{sec:timings}

The program {\tt FastJet}\footnote{Available from
  \url{http://www.lpthe.jussieu.fr/~salam/fastjet}.} %
codes concrete implementations of the $N^2$ and $N \ln N$ algorithms
described above. It has been written in C++ and for the $N \ln N$ case
makes use of a number of pre-existing components.
%
%{\tt FastJet}, the concrete implementation of the code carrying out
%$k_t$ jet-finding in $N\log N$ time, has been written in C++ making
%use as much as possible of pre-existing components.
%
Construction of a Voronoi diagram is a sufficiently common task
(useful in areas of science ranging from astronomy to zoology) that
several codes are publicly available. Of these, the only one that we
found that also straightforwardly allows the addition and removal of
points from a pre-constructed Voronoi diagram, was the Computational
Geometry Algorithms Library (CGAL) \cite{CGAL}, in particular its
triangulation components \cite{CGALTriang}.\footnote{ One issue
  relates to the fact that we need nearest-neighbour location on a
  cylinder ($\eta$-$\phi$ space) whereas CGAL works on the Euclidean
  plane. This problem can solved by making mirror copies of a small
  ($\sim 1/\sqrt{N}$) fraction of the vertices near the $0-2\pi$
  border.} %
For the binary tree we made use of a (red-black) balanced
tree.\footnote{Balanced trees are the basis of the \texttt{map} and
  \texttt{set} classes in the C++ Standard Template Library.}

\begin{figure}[t!]
\begin{center}
\includegraphics[width=\TimingsWidth]{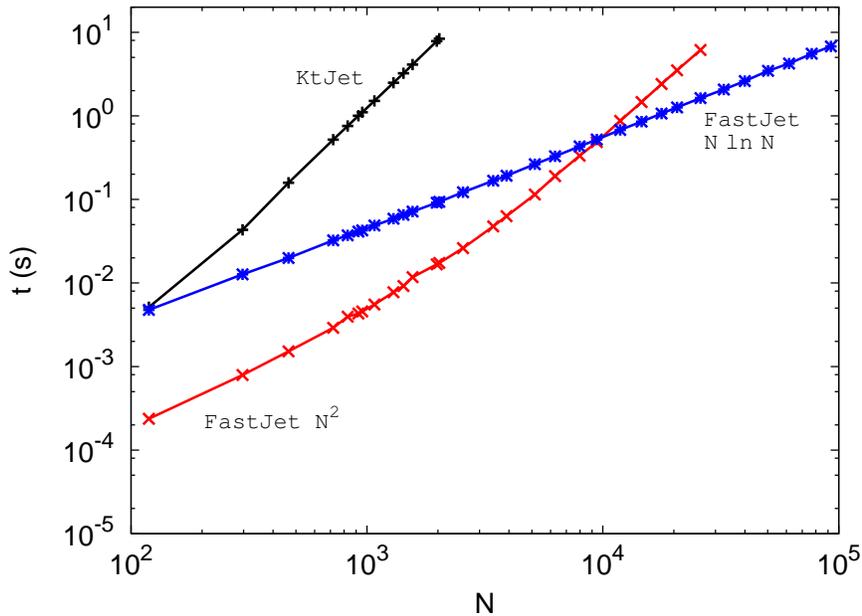}
\vspace{.1cm}
\caption{\label{fig:timings} \small The running times (on a 3 GHz Pentium 4
  processor with $1$~GB of memory, 512~kB of cache, and version 3.4 of
  the GNU {\tt g++} compiler) of the 
  KtJet~\protect\cite{KtImplementation} and FastJet implementations of
  the $k_t$-clustering jet-finder versus the number of initial
  particles. Different values of $N$ have been obtained by taking a
  LHC dijet event with $p_t \simeq 60$~GeV and adding on
  variable numbers of minimum bias events. Both kinds of events have
  been simulated with Pythia~6.3 \protect\cite{Pythia}.}
\end{center}
\end{figure}

Figure~\ref{fig:timings} shows the running times for the two algorithms
in {\tt FastJet} as well as for \texttt{KtJet}, a standard
implementation \cite{KtImplementation} of the $N^3$ algorithm.
%
%%Figure~\ref{fig:timings} shows the running times of our {\tt FastJet}
%%implementation
%%of the two efficient algorithms described in this Letter. The
%%\texttt{KtJet} curve represents a standard implementation
%%\cite{KtImplementation} of the $N^3$ algorithm.
%
Our ``$N^2$ algorithm'' actually departs slightly from exact $N^2$
behaviour owing to certain further optimisations carried
out.\footnote{The coefficient of $N^2$ can be reduced by tiling the
  plane into rectangles of edge length $\ge 1$. Then for each vertex
  $i$ one can limit the nearest neighbour search to its own tile and
  adjacent tiles --- vertices further away will have $R_{ij} > 1$ and so
  $d_{ij} > d_{iB}$.}  The scaling with $N$ of the Voronoi-based
algorithm has been verified to go as $N \ln N$, as expected.
It is the fastest algorithm only for $N \gtrsim 10^4$, owing to a
large coefficient in front of its $N \ln N$ behaviour, mostly
associated with the computational geometry tasks.  This situation
could conceivably be improved in the future by optimisations of the
CGAL package or by replacing it with a dedicated implementation of the
construction and updating of the Voronoi diagram.

The better of the $N^2$ and $N\ln N$ algorithms (which can be selected
based on the value of $N$) runs at least an order of magnitude faster
than the $N^3$ algorithm for all values of $N$ shown, vastly more at
large $N$.

\begin{figure}[t!]
\begin{center}
\includegraphics[width=\TimingsWidth]{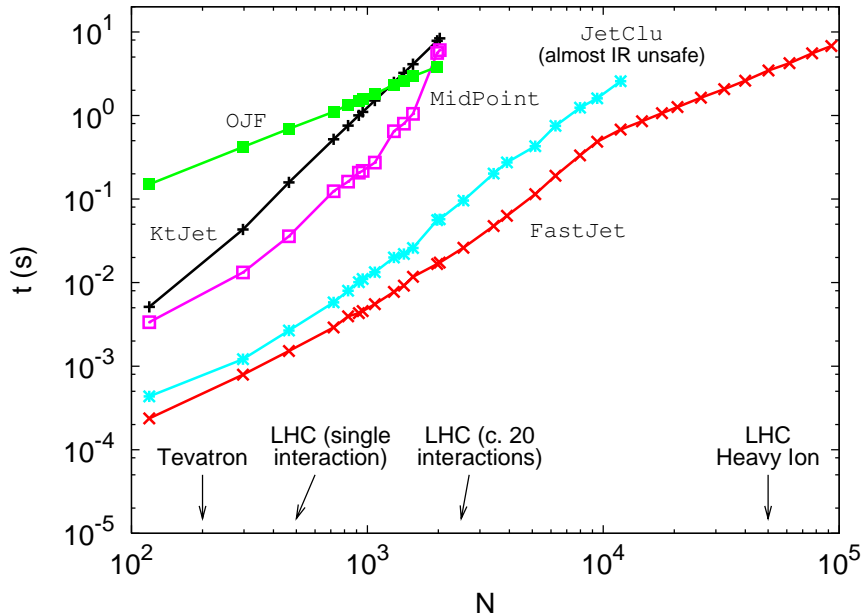}
\caption{\label{fig:comparisons} \small The running times of various
  jet-finders versus the number of initial particles. JetClu is a
  widely-used cone-type jet-finder, however it is `almost infrared
  unsafe', \ie perturbative predictions have large logarithmic
  dependence on small parameters (\eg seed threshold)
  \cite{MidpointIdea,Seymour:1997kj}.
  MidPoint \cite{MidpointIdea} is an infrared safe cone-type jet
  finder. For both we use code and parameters from CDF~\cite{CDFCones}.
  The Optimal Jet Finder~\protect\cite{Tkachov} (OJF) has been run
  with $\Omega_{cut} = 0.15$ and a maximum of 8 jets, so as to produce
  a final state similar to that returned by the $k_t$ and cone
  jet-finders and to limit its run time.
%% When running without the limit on the number of jets run times
%% for even a few hundred particles blew up to tens of seconds
%% (that is with max njets = 40)
}
\end{center}
\end{figure}

Figure~\ref{fig:comparisons} compares the running time of our combined
$N^2$-$N\ln N$ \texttt{FastJet} implementation of the $k_t$ jet-finder
with other jet-finders whose code is publicly available. One sees that
it runs at least an order of magnitude faster than all others (except
the almost IR unsafe JetClu).

\section{Perspectives}
\label{sec:perspectives}

Since the {\tt FastJet} algorithm is functionally equivalent to the
standard $N^3$ algorithms used for the $k_t$ jet finder till now, the
results of the clustering are of course identical to those of other
implementations.
Howewer, its enhanced speed opens up new ways of using
$k_t$ clustering in the analysis of high-multiplicity events. 

Historically, one apparent drawback of $k_t$-type jets with respect to
cone-type jets in hadron-hadron collisions was considered to be the
larger fluctuations of the areas of the jets defined by the clustering
procedure. Such fluctuations would seem to make it more difficult to
subtract, from the hard event, the energy coming from the
non-perturbative underlying event and from any additional minimum bias
interactions taking place in the same bunch crossing (pileup).

However, the fluctuations become irrelevant if one can easily estimate
the area of {\sl each individual} jet. This can be done on an
event-by-event basis, as follows: because of the infrared safety of
the $k_t$ jet-finder algorithm, one can add a large number of
extremely soft particles (``ghosts'') to the event without modifying
the properties of the hard jets. After clustering, each jet will
contain a subset of the ghosts, and if the ghosts had a uniform
density in $\eta$ and $\phi$, then the number of ghosts in a given jet
will be a measure of its area. In practice we have found that the use
of $\order{10^4}$ ghost particles is necessary to obtain reliable area
estimations.  This definition for the area of a $k_t$ jet can of
course be implemented with any coding of the jet-finder. It is however
impractical, indeed nearly impossible, to deal with the required large
number of ghost particles without a fast code.

Preliminary studies have shown that with simple assumptions about the
uniformity of the underlying event and pileup, one can readily
determine its size and subtract it from the hard jets, leading to good
determinations of kinematical quantities (e.g. invariant masses) in
high-luminosity $pp$ collisions, or of single inclusive jet
distributions in Pb-Pb collisions at the LHC. Full results will be
shown in~\cite{cacciarisalam2}.

Two more observations are worth making before closing this section. They
will both be discussed in more detail in~\cite{cacciarisalam2}.

The first is that it can also be interesting to examine alternative
definitions
of jet areas. One option is to make use of the areas of the Voronoi
cells of all the real particles belonging to a given jet. This
definition avoids the need to cluster thousands of ghost particles
together with the real ones. It instead rests on the geometrical
properties of the event, and on the computational geometry component
of the {\tt FastJet} implementation.

The second observation is that there exist clustering-type jet-finders 
other than the $k_t$ jet-finder that share a large fraction of its
features (including, of course, infrared safety), and the possibility
of a fast implementation. One such example is the ``Cambridge''
jet-finder. It was originally formulated in the context  of $e^+e^-$
collisions in~\cite{Dokshitzer:1997in} and an inclusive version,
adapted to hadron collisions, was given in~\cite{Wobisch:1998wt}. We
shall call this inclusive version the Cambridge/Aachen algorithm. It
is defined in the same way as the
$k_t$ jet-finder at the beginning of Section~\ref{sec:kt}, but
replacing the particle-particle distance by $d_{ij} = R_{ij}^2/R^2$,
and the particle-beam distance by $d_{iB} = 1$. We shall show
in~\cite{cacciarisalam2} that the Cambridge/Aachen jet-finder has
smaller average areas than the $k_t$ jet-finder, making it 
perhaps even better suited for jet studies in
high-multiplicity environments.

\section{Conclusions}
\label{sec:concl}

To conclude, we have identified an unexpected relation between
clustering type jet-finders and a widely studied problem in
computational geometry. The resulting reduction of the complexity of
the $k_t$ jet-finding problem, from $N^3$ to $N \ln N$, opens up the
previously inconceivable option of using the $k_t$ jet-finder for the
large values of $N$ that arise when considering all cells of a finely
segmented calorimeter and for heavy-ion events. For moderate $N$, the
one or two orders of magnitude in speed that we gain with a related
$N^2$ approach pave the way to much wider use of the $k_t$ jet finder
for normal hadron-collider jet studies, especially at the LHC.
More generally, the speed gains discussed in this paper also suggest
novel ways of using the $k_t$ jet finder, which are the subject of
ongoing investigation. One example, given in
section~\ref{sec:perspectives}, is the determination of jet areas,
knowledge of which is crucial for optimal subtraction of pileup
contamination in high luminosity environments.

%======================================================================
%\section*{Acknowledgements}
\acknowledgements

We wish to thank J.M.~Butterworth, J.~Huston, T.~Kluge, C.~Roland and
M.~H.~Seymour and discussions and comments. This work was supported in
part by grant ANR-05-JCJC-0046-01 from the French Agence Nationale de
la Recherche.

%======================================================================

%%% Local Variables:
%%% mode: latex
%%% TeX-master: "fastjet-plb.tex"

\end{document}